World Scientific
www.worldscientific.com

# Multi-Band Acoustic Monitoring of Aerial Signatures

Andrew Mead[*,¶], Sarah Little[*,†], Paul Sail[*], Michelle Tu[*,‡],
Wesley Andrés Watters[*,‡], Abigail White[*,§] and Richard Cloete[*]

[*]Galileo Project, Harvard-Smithsonian Center for Astrophysics
60 Garden Street, Cambridge, MA 02139, USA

[†]Scientific Coalition for UAP Studies
Fort Myers, FL 33913, USA

[‡]Whitin Observatory, Wellesley College
106 Central Street, Wellesley, MA 02481, USA

[§]Harvard University
60 Garden Street, Cambridge, MA, USA 02138
[¶]andrew.mead@cfa.harvard.edu

The acoustic monitoring, omni-directional system (AMOS) in the Galileo Project is a passive, multi-band, field microphone suite designed to aid in the detection and characterization of aerial phenomena. Acoustic monitoring augments the Project's electromagnetic sensors suite by providing a relatively independent physical signal modality with which to validate the identification of known phenomena and to more fully characterize detected objects. The AMOS system spans infrasonic frequencies down to 0.05 Hz, all of audible, and ultrasonic frequencies up to 190 kHz. It uses three distinct systems with overlapping bandwidths: infrasonic (0.05 Hz − 20 Hz), audible (10 Hz − 20 kHz), and ultrasonic (16 kHz − 190 kHz). The sensors and their capture devices allow AMOS to monitor and characterize the tremendous range of sounds produced by natural and human-made aerial phenomena, and to encompass possible acoustic characteristics of novel sources.

Sound signals from aerial objects can be captured and classified with a single microphone under the following conditions: the sound reaches the sensor; the sound level is above ambient noise; and the signal has not been excessively distorted by the transmission path. A preliminary examination of the signal and noise environment required for the detection and characterization of aerial objects, based on theoretical and empirical equations for sound attenuation in air, finds that moderately loud audible sources (100 dB) at 1 km are detectable, especially for frequencies below 1 kHz and in quiet, rural environments. Infrasonic sources are detectable at much longer distances and ultrasonic at much shorter distances.

Preliminary aircraft recordings collected using the single, omni-directional audible microphone are presented, along with basic spectral analysis. Such data will be used in conjunction with flight transponder data to develop algorithms and corresponding software for quickly identifying known aircraft and characterizing the sound transmission path.

Future work will include multi-sensor audible and infrasonic arrays for sound localization; analysis of larger and more diverse data sets; and exploration of machine learning and artificial intelligence integration for the detection and identification of many more types of known phenomena in all three frequency bands.

---

[¶]Corresponding author.

## 1. Introduction

The Galileo Project (the Project) seeks to distinguish between human, natural, and scientifically anomalous aerial phenomena using ground-based instruments that utilize passive remote sensing modalities (Watters, 2023). Acoustics offer such a signal, one that is generated and transmitted by a physical process fundamentally different from electromagnetic (EM) waves. Sound travels as longitudinal pressure waves that transmit mechanical energy through a material medium. These waves create variations in sound pressure level (SPL) that are measured in units of Pascal (Pa), where atmospheric pressure at sea level is about 101 kPa. Objects in the Earth's atmosphere can generate sound either intrinsically, like a jet engine (Morinaga *et al.*, 2019) or fireworks; extrinsically, through an object's interaction of the surrounding medium, like wing-tip vortices (Bradley *et al.*, 2007) and sonic booms (Brown *et al.*, 2007); or both, for motor-driven rotor craft like drones (Alexander & Whelchel, 2019) and helicopters (Wang, 2002), or a meteoric explosion (Ens *et al.*, 2012).

The inclusion of an acoustic monitoring, omni-directional system (AMOS) in Phase 1 of the Project serves as an extra layer of validation when categorizing visual observations as known or unknown. For example, planes, helicopters, and drones all have identifiable audio signatures. The simultaneous measurement of sound waves and visual observations will enable detections, identifications, and characterizations to be compared between independent sensor systems. Further, AMOS may be able to detect, identify, and characterize targets in environmental conditions where the detection of EM signatures is challenging (e.g. water vapor, not in line-of-sight).

In the simple model of sound propagating through a volume of homogeneous atmosphere in free space, sound attenuation occurs through geometric divergence and atmospheric absorption. Geometric divergence is simply the spherical spreading of a sound wave from a point source, and results in an SPL decrease of 6 dB per doubling of distance from the source, independent of frequency. Atmospheric absorption, on the other hand, depends on frequency, humidity, temperature, pressure, and composition, and is related to the relaxation frequencies associated with the vibration of nitrogen and oxygen molecules in air (Kapoor *et al.*, 2018). The atmospheric absorption coefficients (decibels per meter) in the Earth's atmosphere, calculated as a function of frequency, increase as frequency rises. Thus, the energy of low-frequency waves can propagate long distances, while at high frequencies, the energy is quickly absorbed and travels only short distances.

Further complicating the natural acoustic environment, the atmosphere is not homogeneous. Sound speed in the Earth's atmosphere varies with altitude (Bass *et al.*, 2007). These layered sound speed variations can cause reflections and ducting of energy into sound channels, which in turn can increase or decrease travel distance. Wind speed and direction, which are also functions of altitude (White & Hoffman, 2021), affect the apparent sound speed as recorded by ground observers, carry sound longer or shorter distances than predicted by geometric divergence and absorption, and create turbulent eddies which distort and attenuate sound transmission (Stubbs *et al.*, 2005).

Geophysical sources in the environment can generate the entire spectrum of sound that is transmissible in the Earth's atmosphere, from the low-frequency limit around 0.003 Hz, which separates sound from atmospheric gravity waves (Bittner *et al.*, 2010), to a practical high frequency limit around 1 MHz, where sound in air cannot propagate more than a meter (NPLb, 2022; Vladišauskas & Jakevičius, 2004). Such natural sound sources include wind, rain, storms, lightning, earthquakes, volcanic eruptions, methane release, geysers, landslides, avalanches, tree falls, ocean waves, waterfalls, meteors, bolides, and auroras.

Audible sound is defined by the frequency range of human hearing and runs from 20 Hz to 20 kHz. Using the speed of sound in dry, 20°C air at sea level as 343 m/s (NPLa, 2022), these audible frequencies correspond to wavelengths ranging from approximately 17 m to 0.017 m. Acoustic frequencies below 20 Hz are defined as infrasound, and frequencies above 20 kHz are defined as ultrasound. The AMOS audio system will record sounds in all three bands, from 0.05 Hz to 190 kHz, corresponding to wavelengths ranging from 6.86 km to 1.8 mm.

Animals also generate a wide range of sounds in the environment, predominantly in the audible band. Some animals, such as elephants and emus, produce infrasound. Elephant communication has been recorded in the 14–24 Hz range (Payne *et al.*, 1986). Other animals, such as bats, moths, and katydids, produce ultrasound. The highest known pure tone from an animal is produced by a trident bat, *Cleotis percivali*, a native of Southern Africa, that emits calls with a carrier frequency of 212 kHz (Thiagavel *et al.*, 2017).

Much of human machinery generates sound in the audible range, although not by design, e.g. engines, motors, jets, rotors, wheels, jack-hammers, compressors, generators, leaf blowers, fans, etc. In the infrasonic range, human sources include wind turbines, trains, supersonic aircraft, bombs, explosions, and demolition. In the ultrasound range, human-made devices generate signals both intentionally, e.g. proximity and motion detectors, and unintentionally, as a byproduct of some electronic components.

Despite the complex environmental and ambient soundscape of the Earth's atmosphere, many animals successfully use sound to communicate, navigate, find and catch prey, and avoid predators. Bats use echolocation in complex acoustic environments, and elephants communicate over hundreds of kilometers by taking advantage of the propagation distance of low frequency sound.

Acoustic sensors have been adapted by scientists for a wide variety of applications. Field biologists use remote acoustic classification and monitoring to study such diverse creatures as bats (Frick, 2013), beehives (Kulyukin *et al.*, 2018), birds (Stowell *et al.*, 2016), frogs (Xie *et al.*, 2016), and cane toads (Dang *et al.*, 2008). Atmospheric researchers use ground-based infrasound arrays to help detect, locate, and characterize meteors (ElGabry *et al.*, 2017), bolides (Ens *et al.*, 2012), auroras (Wilson, 2005), and tornadoes (Elbing *et al.*, 2019).

Engineers have also adopted sophisticated acoustic sensor arrays and processing algorithms to solve problems related to the detection and identification of objects in the sky. Applications range from separating aircraft sound from ambient noise (Klaczynski, 2014), tracking light-weight aircraft via propeller noise (Tong *et al.*, 2013), detecting and identifying low-flying planes (Sutin *et al.*, 2013), detecting illegal drones (Sedunov *et al.*, 2019), to gathering intelligence on long-distance battlefield equipment (Stubbs *et al.*, 2005).

Most of these applications involve multi-microphone arrays to allow for target localization, noise reduction, and signal enhancement. The future goal of the Project's acoustics program is to implement such features, but for this paper we consider the capabilities and limitations of object detection, identification, and characterization using single, omni-directional microphones. The sound frequency spectra of known objects extends well beyond both ends of the audible frequency spectrum. These bands will have their own unique yet overlapping sound sources. To span such a wide frequency range, AMOS includes three independent sensor systems, each with its own microphone, optimized for three different bandwidths: infrasonic, audible, and ultrasonic.

The role that acoustics will play in achieving the Galileo Project's objectives involves both supporting and leading elements. It will provide an additional sensor modality with which to detect and identify aerial objects in real-time. In the future, it may be able to support localization and tracking by providing approximate coordinates for targets that are obscured by clouds. Acoustics will lead in providing direct information about the acoustic source characteristics of phenomena. This will include distinguishing between known propulsion methods (e.g. jet, rocket, propeller, rotor), and gaining information about the near-field interactions of objects with the surrounding atmosphere (e.g. shock waves, turbulence, vortices, resonant structures). Further, if the transmission path is well characterized, acoustics has the potential to determine shock wave source location, structural vibration characteristics, and detailed propulsion-related or other characterizing acoustic emissions that may or may not have any associated EM signature.

According to witness reports, the connection between unidentified aerospace phenomena (UAP) and sound is ambiguous. Analysis of historical UAP report databases have found that many witnesses report no sound associated with UAP, while a smaller subset include witness reports of hearing whistling, hissing, rumbling (Johnson & Saunders, 2002), and buzzing, humming, beeping, or pulsing (UKMoD, 2000). Some reports explicitly note an absence of sound when sound was expected.

It should be noted here that the acoustic sensitivity of human hearing is not flat across the frequency spectrum. It is most sensitive to frequencies from 1 kHz to 6 kHz, and less sensitive to higher and lower frequencies (AE, 2022). In contrast, AMOS's audible microphone has a relatively flat frequency response from 3.5 Hz to 20 kHz. Therefore, the AMOS system will be sensitive to a wider bandwidth of frequencies than found in witness reports of sound.

Even in cases where no discernible sound is reported, it cannot be assumed that the object was not producing sound. For example, the observer may have been too far away to hear the sound, given attenuation and damping relative to the ambient noise floor. Likewise, sound energy can be outside the range of human hearing, which would manifest in either infrasonic ($< 20\,\text{Hz}$) or ultrasonic ($> 20\,\text{kHz}$) frequencies. Under certain atmospheric conditions, sound generated in the upper atmosphere, including sonic booms, can be refracted upwards, never reaching the ground (Kästner & Heimann, 2010).

Historical UAP witness reports have also described reactions by animals, suggestive of ultrasonic or infrasonic frequencies to which some animals are sensitive. In considering animal reactions, it may be the cessation of ambient animal noise that is notable. These considerations underscore the importance of monitoring sound across an exceptionally broad frequency spectrum.

The shortage of peer-reviewed publications describing efforts to identify UAP through field instrumentation is discussed in Watters (2023). In addition, details are provided of the instrumentation suites for six well-documented scientific field studies: Hessdalen Valley, Norway (Strand, 1984; Teodorani & Strand, 2001; Teodorani, 2004); Ontario, Canada (Teodorani, 2009); Yakima Valley, WA (Akers, 1972, 1974, 2001, 2007); Piedmont, MO (Rutledge, 1981); Würzburg, Germany (Vodopivec & Kayal, 2018; Kayal, 2020); and French airspace (SIMGA2, 2021; Colas et al., 2020). All the systems described included optical cameras, and many had radio, spectroscopic, and magnetic field instruments. The Yakima Valley and Piedmont studies were outfitted with ultrasonic microphones, but no recorded acoustic data are presented in any of their reports. One seven-day infrasonic study at Hessdalen (Farges et al., 2011a,b) was beset by noise during the day due to high winds, and produced no nighttime infrasonic detections that correlated with the six optical detections.

An article published in the Journal of the British Interplanetary Society (Stride, 2001) fully details an experimental methodology and the design of a long-term, multi-sensor, passive autonomous data acquisition system for use in the search for extraterrestrial vehicles (SETV). This design included an ultrasonic microphone, but the system was never developed or fielded.

Related camera-based observational systems include six fireball and meteor tracking networks that are currently in use around the world; the details of their operational characteristics are described in Szenher (2023). Detections from these networks can be correlated with the Incorporated Research Institutions for Seismology (IRIS, 2022) global infrasound network to assist in characterizing meteors.

In short, no publicly available data on sound has been generated from previous or ongoing scientific attempts to identify anomalous aerial objects. While various audio recordings of purported UAP acoustic signatures can be found on the internet, the lack of information around their provenance, chain of custody, audio capture methodology, calibration, and corroborating and contextual evidence ensure that audio analysis of such recordings will never yield usable scientific results.

The methodologies described in this paper for detecting, identifying, and characterizing acoustic aerial signatures go far beyond previous efforts in the field. By including infrasonic, audible, and ultrasonic microphones and band-appropriate analyses, this work seeks to create a framework for a comprehensive scientific study of sound in relation to aerial phenomena.

The first steps for AMOS data analysis include determining the signal-to-noise ratio for local ambient conditions and common aerial objects, identifying bandwidths of interest for classes of phenomena, and designing data reduction methods and analysis techniques for the rapid identification of known objects, starting with aircraft of all types. Future goals include expanding the current single-microphone audible system into an array of four microphones to allow for real-time, 3D sound source localization and tracking. The infrasonic system, which detects acoustic wavelengths greater than $17\,\text{m}$ from distant sound sources, will also be expanded and formed into a local array of three sensors for source discrimination and localization. This array will then be calibrated by and joined with data from existing seismic and infrasonic networks such as the United States Geological Survey (USGS, 2022) seismic networks, and the IRIS infrasonic systems, using its jAmaSeis software (SAGE, 2022), for ongoing analysis. The ultrasound system can be expanded into a localization and tracking array if it is determined in Phase 1 that ultrasound plays an important role in object characterization.

This paper provides details on the Phase 1 acoustic instrument suite of the Galileo Project. Section 2 presents the technical specifications of the AMOS instruments, including the infrasonic, audible, and ultrasonic systems, as well as the airplane transponder receiver. Section 3 describes the field installation of the microphones and data collection hardware, including the outdoor tower assembly. Section 4 discusses calibration and ambient noise background considerations for the different sensor systems. Section 5 explores the various physical and environmental limitations associated with outdoor field measurements of sound. Section 6 explores potential approaches to data analysis, and includes some sample data and preliminary analysis.

## 2. Instrumentation

The acoustic suite described in this paper consists of one each of an infrasonic, audible, and ultrasonic sensor system (Fig. 1). These sensors were chosen to span the broadest range of frequencies while using existing off-the-shelf technology. Each sensor chosen was built specifically for measuring sound in the outdoor environment in the bandwidth of interest. For example, the infrasonic sensor is used for monitoring wind turbines, explosions, and meteors. The audible mic was designed to be placed in the ground to monitor aircraft noise near airports, and the ultrasonic sensor is a field-monitoring instrument for bat populations. This section gives the specifications for each instrument sensor and associated data capture device.

### 2.1. Infrasonic System (InfraS)

The primary sensor of the InfraS is the Infiltec™ INFRA20 infrasound monitor. This microbarograph has a solid-state differential pressure sensor and a high-pass pneumatic filter (0.01 Hz). The resolution is 0.001 Pascals (equal to 0.01 microbar or 0.0075 millitorr) over the range of ±25 Pascals (250 microbars or 187.5 millitorr). The frequency response runs from 0.05 Hz to 20 Hz with an eight-pole analog low-pass elliptic filter at 20 Hz. All data for the InfraS is captured via a Raspberry Pi 4 (2 GB) running jAmaSeis (IRIS, 2022) software. The final output format is 16-bit Seismic Analysis Code (SAC) files captured at a sample rate of 50 Hz, which equates to 10 MB of data per 24-h period.

The InfraS will run continuously with real-time and historical data available remotely via the jAmaSeis software and network, with the option to export and save any portions of interest as stand-alone SAC files. Given the relatively small file sizes, no additional local storage is needed for the capture device. The InfraS will regularly interface with a Network Time Protocol (NTP) server in order to minimize clock errors to milliseconds. Data backup of the master SAC files will happen on a monthly basis.

### 2.2. Audible system (AudS)

The primary sensor of the AudS system is an omnidirectional industrial grade GRAS® 41AC-3 outdoor microphone with a relatively flat frequency response from 3 Hz to 20 kHz, configured for 0° of incidence for vertical mounting. The set sensitivity at 250 Hz (±3 dB) is 50 mV/Pa and it has a dynamic range of 26–138 dB or 0.4 mPa–158.9 Pa. This microphone is designed for permanent outdoor installation and is typically used for in-field monitoring of overhead aircraft noise. The 41AC-3 is fully waterproof and fulfills the requirements of Ingress Protection (IP) rating IP55 in the International Electrotechnical Commission (IEC) standards for fully waterproof and dustproof enclosures.

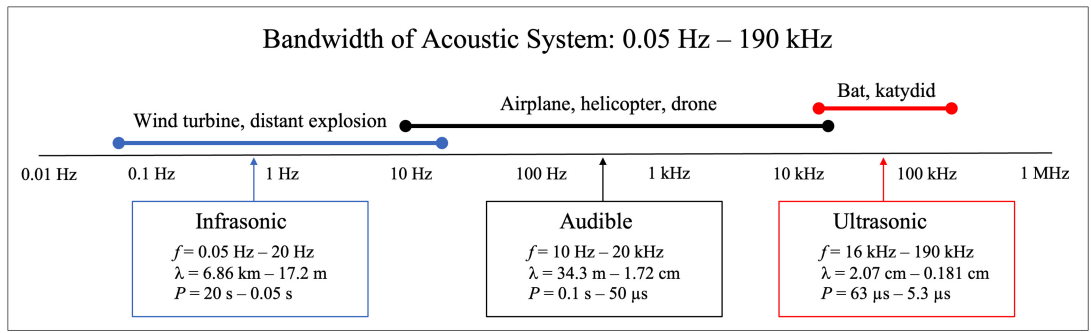

Fig. 1. (Color online) The bandwidths of AMOS's three sensors systems spanning infrasonic, audible and ultrasonic ranges for a total bandwidth of 0.05 Hz to 190 kHz; where $\lambda$ is wavelength (m), $f$ is frequency (Hz), $P$ is period (s), $\lambda = c/f$, and $P = 1/f$, and using a speed of sound in air of $c = 331$ m/s. The generalized frequency ranges of several natural and anthropogenic sound sources are also shown.

The AudS capture device is a Raspberry Pi® 4 (4 GB) with the HiFiBerry® digital-to-analog converter (DAC) and analog-to-digital converter (ADC) Pro installed via hardware-attached-on-top (HAT) specification. The 41AC-3 connects to the ADC via a custom Bayonet Neill–Concelman (BNC) — 1" adapter with an in-line two-pole low-pass filter at 22 kHz for anti-aliasing. The ADC has a frequency response of 10 Hz–20 kHz and captures data as 24-bit mono WAV files with an industry-standard sample rate of 44.1 kHz, which equates to 11.43 GB of data per 24-h capture. The AudS system captures waveform audio file (WAV) data with embedded timecode via NTP in increments of 60 min via the open-source software GStreamer multimedia framework (Cloete, 2023). These data will be recorded to a local external hard drive (2 TB), which will be transferred to long-term storage and locally purged every 10 weeks in order to maintain enough space for ongoing capture.

### 2.3. *Ultrasonic system (UltraS)*

The primary sensor of the UltraS system is the Wildlife Acoustics® SMM-U2 ultrasonic microphone with a calibrated frequency response between 10 kHz – 190 kHz. The SMM-U2 has a built-in two-pole high-pass filter at 1 kHz and uses a cardioid directional pattern configured for vertical monitoring with the null pointed down. The SMM-U2 is fully waterproof and fulfills the requirements of IP68 in ICE's standards for dustproof and waterproof enclosures. The UltraS data capture device is the Wildlife Acoustics® SM4BAT ultrasonic recorder. The SM4BAT has a frequency response of 10 kHz – 190 kHz and includes a selectable (10 kHz or 16 kHz) two-pole high-pass filter. We set the filter at 16 kHz, the lower end of the relatively flat portion of the SMM-U2's frequency response curve. The SM4BAT is enclosed in waterproof polycarbonate housing. The final output format is 16-bit mono WAV files captured at a sample rate of 512 kHz, which equates to 88.40 GB of data per 24-h period.

The SM4BAT will run continuously and will save data locally onto two 2 TB Secure Digital flash memory cards (SDXC). Those cards will need to be manually removed and their data transferred over to long-term storage every six weeks in order to maintain enough space for ongoing capture. The system uses four interchangeable SDXC cards to minimize downtime during data transfers. The system uses an external Global Positioning System (GPS) unit to maintain time. This time-keeping approach differs from the NTP solution for InfraS and AudS because the UltraS is a closed and proprietary system without a straightforward option to integrate with NTP. However, the long-term goal is a single integrated clock solution for all three frequency bands. This will be essential for future multi-band, multi-microphone source localization. Until then, frequency crossover between sensors, and an on-site transient tone generator, will be used to track time-slip attributable to multiple clock sources.

### 2.4. *Automatic dependent surveillance–broadcast (ADS-B)*

The instrument suite includes an ADAfruit® software-defined radio (SDR) ADS-B receiver (antenna and Universal Serial Bus (USB) stick) connected to a Raspberry Pi® 4 (4 GB) for logging the flight transponder data of all commercial and private jets, planes, and helicopters (FAA, 2022). The comma separated variable (CSV) file comprises such information as aircraft identification number, location, ground speed, altitude, and time, from each aircraft within range, once per second. Military aircraft within one mile of a lead aircraft are not required to transmit, nor are smaller aircraft required to carry a transponder in certain types of airspace or with prior permission (AOPA, 2022). Aircraft that fly near the Project's development site will provide data used for assigning ground-level sound profiles to explicitly-defined labels (e.g. using spectral features), or implicitly-defined labels (e.g. defined using a machine learning model), thus creating acoustic IDs for each plane type, and possibly for individual planes. These labels can then be used for rapid aircraft identification. Furthermore, the variations in the acoustic signature of the same aircraft at different times of day or in different weather conditions can be used to characterize the variance in the local acoustic transmission path. This will contribute to determining confidence levels in detection, identification, and characterization of sound emissions from aerial objects.

### 3. Installation

All three systems are housed on the same custom-built tower, designed to minimize vibration and air turbulence that could interfere with sensors (Fig. 2). Phase 1 field testing will inform tower redesign and sensor positioning, if necessary. The AMOS tower is located near the Project's optical and infrared (IR) sensor packages in a spot optimized for minimum wind, ambient noise, and ground-sourced noise. A portable GPS unit measures the geographic coordinates of the microphones. The AudS (41AC-3) and UltraS (SMM-U2) microphones are housed near the top of the tower. The InfraS (INFRA20) unit is housed near the bottom of the tower to minimize wind noise. It is enclosed in a temperature-controlled box to minimize pressure fluctuations related to thermal variations. The capture devices for the InfraS and AudS systems are housed together in a waterproof and shielded electronics enclosure alongside their power, data storage, and data connection elements. The ADS-B SDR USB stick is in the same enclosure and thru-wired to its water-proofed antenna on top. The capture device for the UltraS, the SM4BAT, is fully waterproof from the factory and thus can be mounted without additional protection from the elements. The AudS, InfraS and UltraS electronics enclosures are located on the middle of the tower, with both power and ethernet connections.

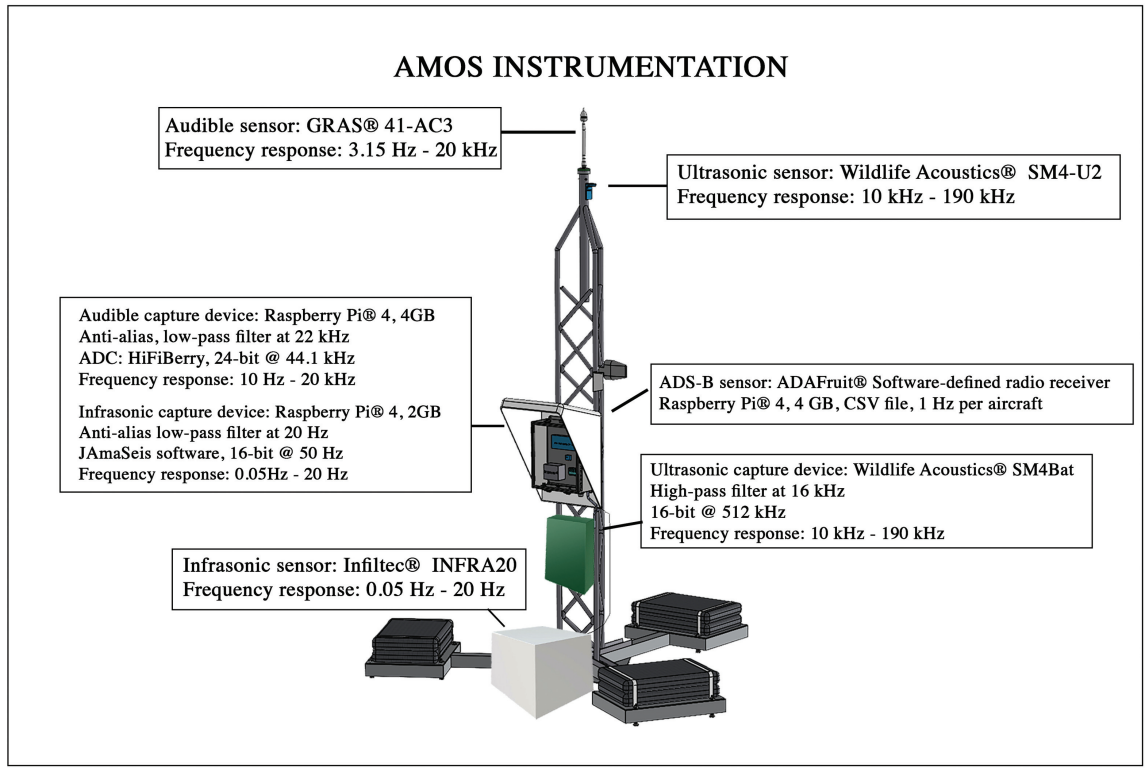

Fig. 2. (Color online) A schematic of the full acoustic sensor suite on its tower, with data collection specifications. Note that the ultrasonic capture device operates with its high-pass filter at 16 kHz, thus constraining the UltraS bandwidth to 16 kHz – 190 kHz; and the audible capture device has a frequency response of 10 Hz–20 kHz, thus constraining the AudS bandwidth.

## 4. Calibration

Each sensor is factory calibrated and shipped with an individual frequency calibration curve. However, the voltage output of each sensor needs to be reca-librated to SPL when a sensor is moved to a new location and integrated into a digital data capture device. This enables us to accurately relate the recorded acoustic power spectrum of aerial objects to their acoustic source SPLs.

The recorded SPL at any given field site will depend on a diverse set of transmission factors such as topography, soil type, vegetation, buildings, weather patterns, temperature, and wind speeds. In addition, sources in the ambient noise field itself will be particular to each site and depend on differing human activities such as those found in urban, suburban, and rural areas, as well as overhead flight traffic, local animal sounds, and sound sources associated with prevailing winds.

Thus, there are three separate types of cali-bration required. The first is field-calibration of the sensors. The second is to characterize the site-specific sound transmission environment by calibrating the frequency response of the acoustic systems to signals of known power, bearing, distance, and frequency. The third is to create a noise benchmark by sampling the ambient sound field, which will be used to set detection confidence thresholds.

Part of the site selection process for the Project's instrument suite will include an estimate of the average noise levels of an area from the U.S. National Transportation Noise Map shown in Fig. 3 (USDOT, 2022). Once a candidate location is selected, a hand-held sound meter will be used on-site to provide sound levels at specific GPS coordinates. Understanding and characterizing the ambient acoustic environment will help to determine our confidence in identifying known objects at each site.

### 4.1. *Sensor SPL and transmission environment calibration*

For AudS, the GRAS® 41AC-3 transducer's frequency response curve is factory calibrated. Once on site, the Tekcoplus® ND9B sound level calibrator will be used to produce a 1 kHz sine wave at 114 dB in close proximity to the GRAS® transducer. The resulting measurements will be checked against the GRAS® 41AC-3 lab calibration data. Any discrepancies in frequency response will be corrected via a filter during data analysis in order to properly relate voltage to SPL.

Fixed sound transmission inhomogeneities in the local environment can also affect the relationship between source SPL and measured acoustic power spectra. To address this, tones of 100 Hz, 500 Hz, and 1 kHz will be played through a speaker at different distances from the GRAS® transducer and the predicted fall-off of SPL (Plotkin *et al.*, 2000) will be compared to actual. This procedure will be carried out at four points of the compass in order to map the underlying directional anisotropy in the sound transmission field that might be due to fixed effects such as topography, buildings, or ground composition. Any directional variations in SPL fall-off need to be considered as noise, adding to signal SPL uncertainty in the omni-directional microphones.

For additional calibration, the ADS-B flight data will allow for correlation of known airplane and helicopter sources with the data recorded by AudS. This will operate as an ongoing system check as well as generate a better understanding of variability in the sound transmission pathways over time.

The UltraS (SMM-U2) will use the Wildlife Acoustics ultrasonic calibrator (SM2CAL) for calibration. The SM2CAL emits a 48 dB, 40 kHz signal when measured at a distance of 30 cm from the SMM-U2. The resulting measurements will be checked against the Wildlife Acoustics SMM-U2 calibration data. Any discrepancies in frequency response will be corrected via a filter during data analysis. The sound source will then be tested at different distances and compass points, and the expected SPL fall-off will be compared with actual.

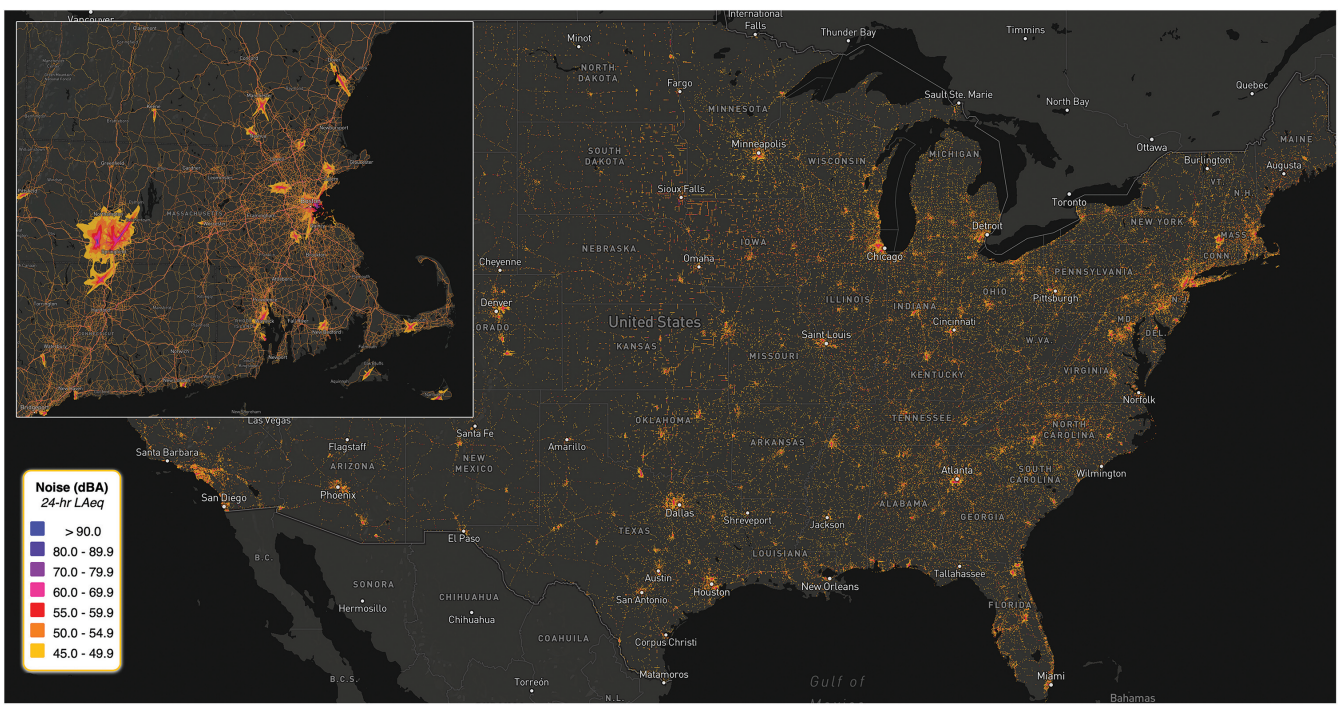

Fig. 3. (Color online) Average aviation, rail, and road noise levels across the United States in 2016 and 2018, as measured by the U. S. Department of Transportation, with a Boston-area insert.

The calibration of the frequency response of acoustic sensors below 100 Hz is not well covered by existing IEC standards, given the challenges around accurately reproducing low-frequency reference tones outside a lab. For this project, the InfraS will be calibrated by data comparison with a network of infrasonic sensors publicly available as well as by comparing the crossover frequency shared with the AudS microphone (GRAS® 41AC-3; 16 Hz–20 kHz).

### 4.2. *Ambient background calibration*

Ambient noise in the environment will be both time and frequency-dependent. Animals, weather, and human activities all have diurnal as well as seasonal components, and there are many other factors that cause variability on different time-scales. Nonetheless, the average ambient noise over a fixed period of days, in dB, is a useful measurement to establish a baseline for initial estimation of detection thresholds for a given site.

The three primary AMOS sensors (INFRA20, 41AC-3, SMM-U2) will be used to determine the on-site, long-term ambient noise benchmarks and their variability. An additional SPL meter, the VLIKE™ sound level meter, which has a measuring scope of 35–130 dB, will establish the initial ambient noise floor readings in dB, providing the expected total loudness, summed over all frequencies, of the acoustic environment as a function of time. Once data collection begins, the AMOS suite will run continuously for roughly seven days and a series of spectral averages will be created and examined to determine noise floor and its variability. The AMOS sensor data will establish hourly average ambient noise patterns to give expected values for noise at different times of day. Detection threshold calculations derived from these benchmarks will be updated on an ongoing basis if changes to the ambient noise field warrant it.

Once the initial average ambient noise floor, spectral response, and variability have been established, estimations can be made to determine the average likelihood of hearing a known sound given its distance to the sensor. The surveys will run until we have characterized the sound transmission and ambient noise variability of the site well enough to get consistent corroboration between the detections from AMOS, the ADS-B data, and the Project's visible and IR cameras.

## 5. Detection Range Estimates

### 5.1. *Sound attenuation*

Fundamental limitations of the AMOS system arise from physical distance constraints in the propagation of sound waves through air due to attenuation. Additionally, local environmental conditions play a key role in determining the likelihood of hearing or not hearing a sound from a distance. Attenuation of sound waves occurs through a variety of mechanisms, all of which combine over distance to decrease SPLs relative to the source level. The International Standards Organization ISO96131 (1996) outlines a general formula for calculating overall attenuation of sound in air:

$$A = A_{\text{div}} + A_{\text{atm}} + A_{\text{gr}} + A_{\text{bar}} + A_{\text{misc}}. \quad (1)$$

This formula gives the overall attenuation $A$, from the geometrical divergence $A_{\text{div}}$, atmospheric absorption $A_{\text{atm}}$, ground effect $A_{\text{gr}}$, barrier effect $A_{\text{bar}}$, and miscellaneous effects $A_{\text{misc}}$, which includes attenuation from foliage, industrial sites, and houses.

For the case of jets and other aerial vehicles, $A_{\text{gr}}$ is replaced by the special case of lateral attenuation $A_{\text{lat}}$, which better fits empirical observations for angles of incidence appropriate for aerial objects and which we will use here. In addition, since barriers, foliage, and houses will have negligible effect on sound from aerial objects above 10 m elevation, the attenuation factors $A_{\text{bar}}$ and $A_{\text{misc}}$ will be set to zero for this analysis. While geometrical divergence and lateral attenuation are independent of frequency, attenuation due to atmospheric absorption is not. In cases where frequency information is not available, ISO96131 (1996) recommends performing the calculation with 500 Hz as the dominant frequency.

Geometric divergence $A_{\text{div}}$ is the attenuation due to the spherical spreading of the sound pressure wavefront from an omni-directional point source radiating into free space. Working in decibels (dB), the formula for the special case where the point source is 1 m from where the reference sound source pressure level is measured, $d_0 = 1$ m, is

$$A_{\text{div}} = 20\log_{10}\left(\frac{d}{d_0}\right) + 11, \qquad (2)$$

where $d$ is the distance from the point source to the receiver.

Atmospheric absorption $A_{\text{atm}}$ is characterized by a frequency-dependent atmospheric absorption coefficient $\alpha$, with units of dB m$^{-1}$, calculated for each frequency and multiplied by the distance between the source and the receiver

$$A_{\text{atm}} = \alpha d. \qquad (3)$$

The atmospheric absorption coefficient is also a function of temperature, relative humidity and, to a lesser extent, pressure. The full equations for calculating the atmospheric absorption coefficients are based on the relaxation frequencies associated with the vibration of nitrogen and oxygen molecules in the atmosphere (ANSI, 1995). As such, these coefficients depend on sound frequency $f$, temperature $T$, air pressure $p$, humidity $h$, and the relaxation frequencies of oxygen $f_{\text{rO}}$, and nitrogen $f_{\text{rN}}$. The empirical formula for this coefficient, $\alpha$, given below, is found in ANSI (1995) and based on ISO96131 (1993). An online calculator provides the same results (MASE, 2022).

The following equation gives the molar concentration of water vapor $h$, where $h_{\text{rel}}$ is relative humidity in percentage, $p_{\text{sat}}$ is saturation vapor pressure, $p_{\text{atm}}$ is the prevailing atmospheric pressure in kilopascals, and $p_{\text{ref}}$ is 101.325 kPa, the reference atmospheric pressure:

$$h = h_{\text{rel}}\left(\frac{p_{\text{sat}}}{p_{\text{ref}}}\right)\left(\frac{p_{\text{atm}}}{p_{\text{ref}}}\right)^{-1}. \qquad (4)$$

The ratio of saturation vapor pressure to the reference atmospheric pressure can be estimated by the following equation:

$$\frac{p_{\text{sat}}}{p_{\text{ref}}} = 10^C, \qquad (5)$$

where exponent $C$ is given by the following equation in which $T_{01}$ is 273.16 K, the triple-point isotherm temperature, and temperature $T$ is in kelvin:

$$C = -6.8346\left(\frac{T_{01}}{T}\right)^{1.261} + 4.6151. \qquad (6)$$

The values of the vibrational relaxation frequencies of oxygen and nitrogen, respectively, are calculated

from the following equations:

$$f_{\text{rO}} = \frac{p_{\text{atm}}}{p_{\text{ref}}}\left[24 + \frac{(4.04 \times 10^4 h)(0.02 + h)}{0.391 + h}\right] \qquad (7)$$

and

$$f_{\text{rN}} = \frac{p_{\text{atm}}}{p_{\text{ref}}}\left(\frac{T}{T_{\text{ref}}}\right)^{-\frac{1}{2}} \times (9 + 280h \times e^{-4.170[(\frac{T}{T_{\text{ref}}})^{-\frac{1}{3}}-1]}). \qquad (8)$$

Using these values, the atmospheric absorption coefficient $\alpha$, in decibels per meter, can be calculated from the following equation, where $T_{\text{ref}}$ is 293.15 K, the reference air temperature:

$$\alpha = 8.686 f^2 \times D, \qquad (9)$$

where

$$D = 1.84 \times 10^{-11}\left(\frac{p_{\text{atm}}}{p_{\text{ref}}}\right)^{-1}\left(\frac{T}{T_{\text{ref}}}\right)^{\frac{1}{2}}$$
$$+ 0.01275 e^{(\frac{-2239.1}{T})}\left(\frac{f_{\text{rO}}}{f_{\text{rO}}^2 + f^2}\right)\left(\frac{T}{T_{\text{ref}}}\right)^{-\frac{5}{2}}$$
$$+ 0.1068 e^{(\frac{-3352.0}{T})}\left(\frac{f_{\text{rN}}}{f_{\text{rN}}^2 + f^2}\right). \qquad (10)$$

The above equations were used to generate Figs. 5–7 in this paper.

The calculation for lateral attenuation $A_{\text{lat}}$ for aircraft sound is slightly different from the ground effect described in ISO 9613. Lateral attenuation is the reduction of sound by the ground relative to angles of incidence, $\beta$, between 0° and 60° (Plotkin et al., 2000). The lateral attenuation model is appropriate for aerial objects and depends on the elevation angle from the sensor to the object.

There are several approaches for calculating lateral attenuation, the most common being the Federal Aviation Administration's Integrated Noise Model (Boeker et al., 2008), which defines the procedure via SAE 1751 (AIR, 1981), and the Department of Defense's NOISEMAP, with its NMAP computational model (Moulton, 1990). Both models use angle of incidence to calculate lateral attenuation. The SAE model predicts attenuations greater than or comparable to the NMAP model (Plotkin et al., 2000). For this reason, we will use the more conservative SAE 1751 guidelines for estimating sound attenuation, which uses the following formula for lateral attenuation of sound from airborne aircraft.

$$A_{\text{lat}} = 3.96 - 0.066\beta + 9.90e^{-0.13\beta} \text{ dB},$$
$$\text{for } 0° \leq \beta \leq 60°, \tag{11}$$

$$A_{\text{lat}} = 0 \text{ dB}, \quad \text{for } 60° < \beta \leq 90°. \tag{12}$$

Values for $A_{\text{lat}}$ are independent of both frequency and distance, with a maximum value of 14 dB at $\beta = 0°$ and a value of 1 dB at $\beta = 45°$.

The total attenuation in our case is the sum of geometrical divergence $A_{\text{div}}$ (dependent on distance), atmospheric attenuation $A_{\text{atm}}$ (dependent on distance and frequency), and lateral attenuation $A_{\text{lat}}$ (dependent on angle of incidence).

With the three dominant sound attenuation factors in hand, the SPL at a receiver $L_{\text{rec}}$, given a known source level $L_{\text{source}}$, is simply the source level minus the total attenuation.

$$L_{\text{rec}} = L_{\text{source}} - A_{\text{div}} - A_{\text{atm}} - A_{\text{lat}}. \tag{13}$$

### 5.2. *The ambient environment*

In order to hear a sound at a distance, the sound levels reaching the receiver must exceed the local environmental noise floor generated by the ambient environment. The power levels of noise in an ambient environment can be measured either as a frequency-independent, total decibel level using a SPL meter, or as a frequency-dependent audio timeseries and converted into the power versus frequency domain through Fourier transforms.

A number of factors may influence the acoustic noise floor, including local traffic; trains; highways; heating, ventilation and air conditioning (HVAC) machinery; wind; running water; and animals. To get some idea of the variability of noise in different environments, we captured sound at three acoustically distinct locations: city, suburban, and rural. A one-minute average of SPL was measured with a sound meter, and a one-minute acoustic sample was extracted from an hour of recording during which the ambient noise remained relatively level. The resulting spectra are displayed in Fig. 4 and show noise levels increasing as we move from rural to city environments, as well as having differing characteristic frequencies.

For a given ambient noise floor $L_{\text{amb}}$, a buffer can be added, $L_{\text{buffer}}$, which represents the number of dB above the noise floor that are necessary to confidently detect the signal. Using Eq. (13), we can determine the source level that can withstand a given attenuation and still remain above the buffer

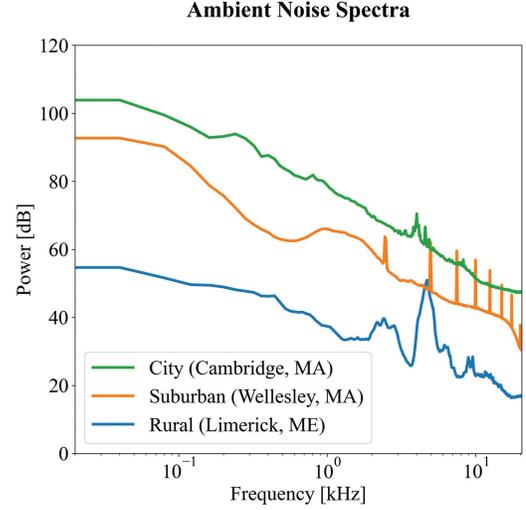

**Ambient Noise Spectra**

Fig. 4. (Color online) Ambient noise spectra of city, suburban, and rural locations derived from a one minute recording in each environment. The average noise floor of the samples and their characteristic peaks are different. The prominent features in the rural spectrum are likely due to natural sounds, including crickets (2 kHz – 8 kHz). In the suburban spectrum, the narrow, high frequency peaks starting at 2.5 kHz are likely due to interference in the recording signal chain and thus were not elements of the acoustic environment. The placement of these harmonics suggests a USB-power supply issue. The broad and irregular peaks in the urban environment are likely due to traffic and HVAC (2 kHz – 6 kHz). However, we were not able to identify the exact sources in this preliminary study.

in the presence of our measured ambient noise level

$$L_{\text{amb}} + L_{\text{buffer}} < L_{\text{source}} - A_{\text{div}} - A_{\text{atm}} - A_{\text{lat}}. \tag{14}$$

Selecting a power level for the detection buffer $L_{\text{buffer}}$ will be critical for automated event detection, and will be part of the ongoing calibration considerations during the system testing and deployment phases. If the buffer level is set too high, events will be missed, and if it is set too low, fluctuations in the ambient noise field will generate false detections. The level of the detection buffer will depend on the consistency of the ambient noise. A level of 10 dB can be used as a starting point, but the exact levels must be determined after the calibration of the system in its ambient environment using sound from known aerial objects. Once a buffer level is chosen, Eq. 14 provides the approximate minimum viable source sound pressure $L_{\text{source}}$ needed to make a detection in the presence of the noise floor at the measurement location. When using a frequency-independent approximation, a dominant frequency of 500 Hz can be assumed, as noted earlier. Using the above formulas, several approximate maximum range estimates can be created for different source power levels. For simplicity in the following calculations, the buffer level $L_{buffer}$ is set to 0 dB.

### 5.3. *Range estimates*

Range estimates depend not only on the ambient noise, but also on the source SPL, otherwise known as its loudness. Table 1 gives some examples of the loudness, in dB, of common sounds in our built environment at a close distance to the source.

#### 5.3.1. *Frequency-independent range calculations*

Using a fixed frequency of 500 Hz, a variety of approximate distance calculations can be made. The following nominal conditions will be used: temperature = 20°C; pressure = 101.325 kPa; relative humidity = 70%. The maximum distance a 500 Hz signal can travel before it falls below a measured ambient noise floor (city, suburban, rural) at varying source levels is shown in Table 2. Based on this approximation and the average conditions above, the sound from a jet would be above ambient noise in a rural setting up to 14.5 km away. Likewise, a normal conversation would be at ambient noise level when 0.4 m away in a noisy urban setting.

Plotted together, the source levels from Table 2 show the decrease in power over distance, ultimately falling below the noise floors of the various environments (Fig. 5). This decrease initially stems from geometric divergence, but over larger distances, the effects of atmospheric absorption are also felt.

#### 5.3.2. *Frequency-dependent range calculations*

Using the above formulas, we can now examine the maximum distance a sound will travel as a function of source frequency. In this case, we fix the source level at 125 dB (as measured at 1 m) and examine the effect of varying the source's frequency (tone) on the distance at which the signal falls below the noise

Table 1. Loudness of common sources of human-made sounds at the given distance from their source. Sound pressure level in decibels.

| Common Examples for Measured dB Levels | | |
|---|---|---|
| Source level (dB) | Measurement distance (m) | Example |
| 140 | 50 | Jet aircraft, threshold of pain |
| 120 | 1 | Loud car horn |
| 110 | 1 | Chainsaw |
| 100 | 1 | Disco in front of speaker |
| 90 | 10 | Diesel truck |
| 80 | 15 | Freight train |
| 70 | 1 | Vacuum cleaner |
| 60 | 1 | Normal conversation volume |

Table 2. Distance at which signal strength drops to the noise floor, for different source levels, at a fixed frequency of 500 Hz. Angle of incidence 45°; ambient conditions: 20°C, 70% humidity, and 101.325 kPa at sea level. Power levels fall off with distance due to geometric divergence, absorption, and lateral attenuation (1 dB). When sound levels drop below the ambient noise levels, the signal cannot be detected. For example, a 60 dB conversation can be heard at 0.4 m in an urban setting, but at 14 m in a rural setting. A 140 dB signal of the same frequency, however, can be heard at 14.5 km in a rural setting.

| | Detection distance for 500 Hz source | | |
|---|---|---|---|
| Source level (dB) | Distance (m) at which signal falls below "city" noise floor of 56 dB | Distance (m) at which signal falls below "suburban" noise floor of 45 dB | Distance (m) at which signal falls below "rural" noise floor of 26 dB |
| 140 | 2,600 | 5,700 | 14,500 |
| 130 | 1,000 | 2,800 | 9,300 |
| 100 | 40 | 130 | 1,100 |
| 80 | 4 | 14 | 130 |
| 60 | 0.4 | 1.4 | 14 |

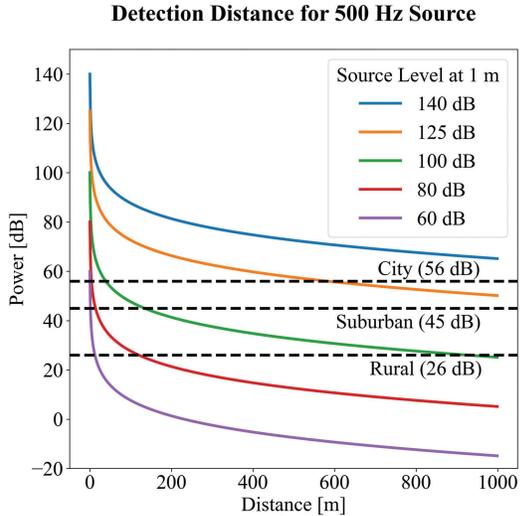

**Detection Distance for 500 Hz Source**

Fig. 5. (Color online) The distance that a 500 Hz signal travels before its power level drops below ambient levels, for different source levels. Angle of incidence 45°; ambient conditions: 20°C, 70% humidity, and 101.325 kPa at sea level. Power levels fall off with distance due to geometric divergence, absorption, and lateral attenuation (1 dB). When sound levels drop below the city, suburban, or rural ambient noise levels, the signal cannot be detected in those locations.

floor, in different ambient environments. Table 3 makes it clear that lower frequencies can travel further; this is due to their lower atmospheric absorption coefficients. Plotting a set of distances (50 m–1 km) relative to frequency attenuation (Fig. 6) shows that signal levels at higher frequencies fall below the noise floors much sooner than for lower frequencies. Low frequencies remain detectable at long distances. Table 3 shows the maximum range at which 125 dB source level sounds of different frequencies remain detectable.

Additionally, we can look at the effect of distance versus frequency more globally to illustrate the dramatic attenuation of high frequency sounds over short distances (Fig. 7). For ultrasonic frequencies near 20 kHz, the maximum distance a 125 dB sound can travel before its power drops to 0 dB is several hundred meters. This distance shrinks with increasing frequency. A 125 dB source at 190 kHz, representing the highest frequency recordable by the AMOS system, would drop to 0 dB in about 15 m. Due to the even closer proximity needed to detect frequencies higher than 190 kHz from an aerial object, and the data handling costs associated with continuous sampling at rates higher than 512 kHz, an ultrasonic microphone for frequencies above 190 kHz will not be used in AMOS at this time.

### 5.4. *Range conclusions*

The environmental noise floor in each frequency band of interest presents a key limiting factor in determining whether a signal will be detected. Once a field site is selected, a background noise survey must be conducted in order to perform signal range calculations as outlined above. Such a survey may also reveal reflective surfaces, nearby barriers, poor

Table 3. Distance at which signal strength drops to the noise floor, for different frequencies, at a fixed source strength of 125 dB at 1 m. Angle of incidence 45°; ambient conditions: 20°C, 70% humidity, and 101.325 kPa at sea level. Power levels fall off with distance due to geometric divergence, absorption, and lateral attenuation (1 dB). When sound levels drop below the ambient noise levels, the signal cannot be detected. For example, a 500 Hz tone can be heard up to 1 km in an urban setting, but up to 9.3 km in a rural setting. A 5 kHz tone of the same power level, however, can only be heard up to 425 m in a rural setting.

| | | Detection distance for 125 dB source | | |
|---|---|---|---|---|
| Frequency (Hz) | Absorption coefficient (db/m) | Distance (m) at which signal falls below "city" noise floor of 56 dB | Distance (m) at which signal falls below "suburban" noise floor of 45 dB | Distance (m) at which signal falls below "rural" noise floor of 26 dB |
| 190,000 | 8.405033 | 5 | 6 | 8 |
| 100,000 | 3.814452 | 9 | 12 | 16 |
| 20,000 | 0.420152 | 54 | 73 | 110 |
| 10,000 | 0.117506 | 125 | 190 | 315 |
| 5,000 | 0.033446 | 255 | 445 | 850 |
| 1,000 | 0.004978 | 500 | 1,200 | 3,400 |
| 500 | 0.002791 | 600 | 1,500 | 4,700 |
| 250 | 0.001124 | 700 | 1,900 | 8,000 |
| 100 | 0.000220 | 700 | 2,300 | 16,000 |
| 20 | 0.000009 | 700 | 2,500 | 22,000 |

**Detection Distance for 125 dB Source**

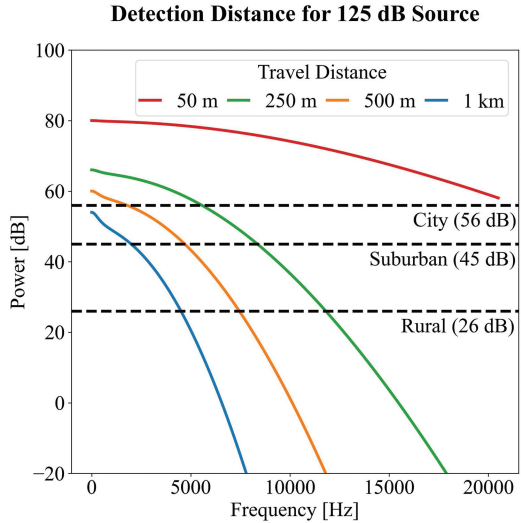

**Power Level Contours for 125 dB Source**

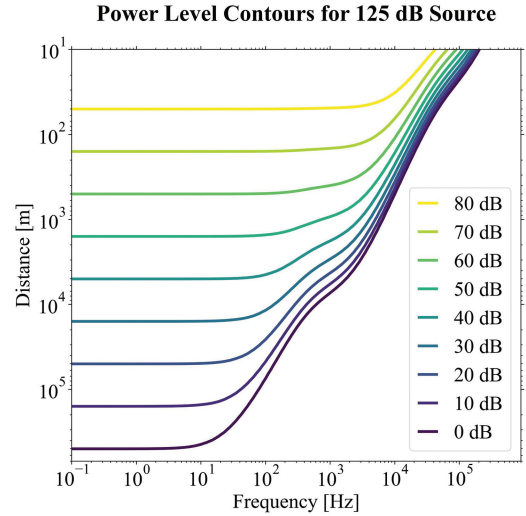

Fig. 6. (Color online) The distance that sound from a source of strength 125 dB at 1 m travels before its power level drops below ambient noise levels, as a function of source frequency in the infrasonic and audible bands. Angle of incidence 60°; ambient conditions: 20°C, 70% humidity, and 101.325 kPa at sea level. Power levels fall off with distance due to geometric divergence and absorption; lateral attenuation is zero. At 50 m, power levels in the infrasonic and audible ranges remain above the noise floor. At 1 km, no frequency has power levels above the city noise floor and only frequencies below 5 kHz will be detectable above a rural noise floor (26 dB).

Fig. 7. (Color online) Sound power level contours as a function of distance and frequency for a fixed sound source level of 125 dB at 1 m, spanning AMOS's entire bandwidth. Attenuation of sound due to geometric divergence and absorption in air, from 10 m (top) to over 100 km (bottom). Angle of incidence 60°; ambient conditions: 20°C, 70% humidity, and 101.325 kPa at sea level. Power levels fall off with distance due to geometric divergence and absorption; lateral attenuation is zero. Infrasonic frequencies of 10–20 Hz travel hundreds of kilometers before their SPL drops to 0 dB, while ultrasonic frequencies of 100 kHz – 200 kHz travel only tens of meters before their SPL drops to 0 dB.

acoustic transmission paths, etc., that may need to be specifically addressed in order to make our detection calculations more accurate.

In addition to being able to detect sound emitted by an aerial object, it is important for AMOS to be able to determine if an object detected by the optical system did not produce sound. With an understanding of sound source levels, range, and frequency versus power attenuation disparities, it may be possible to calculate the likelihood of having "heard" the propulsion sound of an aerial object at a given distance under known conditions, if it existed, and thus to make a determination that an object was soundless. For example, an object that looked like a helicopter in the optical system but made no sound would be potentially anomalous. Understanding expected sound levels will provide insight into whether an object is soundless or not.

Given the immense distances infrasonic sound can travel and the unique aerial acoustic source mechanisms associated with such low-frequencies, for example shockwaves, high-altitude explosions, meteors, lightning, sprites, etc., the infrasonic system has high potential to detect distant and rare phenomena. The ultrasonic system provides good coverage for high frequencies, given the short ranges over which high frequencies can propagate due to atmospheric absorption, and will include detections for known ultrasonic emitters such as bats and insects. The audible system, with a frequency response that overlaps both the infrasonic and ultrasonic systems, anchors AMOS and ensures its ability to accurately monitor an extremely wide bandwidth of sound from aerial objects.

## 6. Data Analysis

### 6.1. *Overview*

Analytical methods for processing the data acquired by AMOS will need to encompass the wide range of sound frequencies and their corresponding atmospheric transmission and environmental noise environments. For instance, infrasonic frequency transmission is affected by regional atmospheric temperature structures, audible frequency transmission by local atmospheric temperature variations, and ultrasonic frequencies by temperatures within a few hundred meters of the sensor. Similar approaches in signal processing and machine learning techniques will be applicable to these three frequency ranges, but each will also require methodologies specific to the spatial scales and atmospheric physics most affecting its acoustic bandwidth.

Initial efforts will focus on analyzing measurements made by single microphones, one in each of the three frequency bands described above. Future work will build upon the wildlife, aircraft, and drone detection solutions referenced in Sec. 1 and incorporate measurements from multi-sensor infrasonic arrays, multi-sensor audible arrays, and correspondingly more complex signal processing algorithms.

The aims of our audio analyses are to extract information related to the detection, identification, and characterization of aerial objects. Audio results will be combined with results from analyses of data collected from the Project's optical, IR and radar instruments, as well as with those of the EM field and particle sensors, in order to detect and classify objects. Additionally, analyses of data in the three frequency bands will be used to characterize sound source mechanisms and relate them to propulsion type, turbulence, resonance, or shock wave generation by the object's movement through the surrounding air.

Analytical approaches will fall into three categories: (1) Traditional analyses, which draw upon the foundation of acoustic digital signal processing (Tohyama *et al.*, 2000) including filtering, noise reduction, spectral analysis, linear predictive modeling, coherence, phase, and time-delay methods; (2) artificial intelligence/machine learning (AI/ML), which draws upon the rapidly developing and highly promising approach of supervised and unsupervised inference models that can be generated with large quantities of data; and (3) a combination of the two approaches, which may prove superior to either one alone. For example, a combined approach was successfully applied to the problem of acoustically monitoring frog population and species richness over large spatio-temporal scales (Xie *et al.*, 2016).

The detection problem is framed with the question, "is there an aerial object within range of our sensors that is making sound?" Using data from a single microphone, we can apply traditional signal processing approaches that increase the signal-to-noise ratio, characterize the ambient noise environment on different time scales, and use time-domain spectral feature enhancement techniques, such as the wavelet transform (Foufoula-Georgiou *et al.*, 1994). The detection problem also requires that AMOS responds to independent detections made by other elements of the Project's instrument suite and make a determination as to whether the object is producing a sound or is silent. For multiple microphones in later stages of this work, signal coherence and time-difference-of-arrival methods will assist with acoustic detection and will add the ability to acoustically localize and track objects (Sathyan *et al.*, 2006).

The identification problem is framed with the question, "is this a sound made by an object already known to the acoustic system?" Solving this problem requires recording sounds from known objects, such as jet aircraft identified with ADS-B transponder data; reducing each complex sound track to a small set of quantifiable characteristics (features); and creating a large database of known features and their corresponding identification labels. Such a database can be rapidly and efficiently searched in order to classify aerial objects into known categories; assign confidence bounds; and highlight and label sounds that are anomalous. This process enables rapid feature extraction from a live data stream and near real-time classification. A simple first approach to acoustic feature extraction is to identify characteristic peak frequencies and compute their ratios. Sounds can also be characterized by their cepstral coefficients, used in speech processing (Beigi, 2011). The cepstrum is the inverse Fourier transform of the logarithm of the power spectrum, and it highlights periodic structures in the frequency domain such as harmonics or reflections of a signal. These approaches reduce the amount of data needed to describe sound from an object by transforming a time series into the frequency domain and identifying characteristic peak frequencies and their relationship to each other. In addition, convolutional neural networks (CNNs) can be used to quantify distinctive features in time-dependent acoustic spectra; CNNs have been used identify models of jet aircraft in this manner (Morinaga *et al.*, 2019). A large labeled sound database can be used to train a CNN, or the features of pre-trained CNNs can be used with semi-supervised learning to build an inference model. Existing labeled sound databases and inference models, such as one for birds (CornellLab, 2022), could also be integrated in the data reduction and analysis pipeline to assist with identification. Similar methodologies are currently being investigated and used in a wide range of research aimed at identifying and monitoring sound-producing animals such as birds, bats, frogs and toads in the field (Rhinehart *et al.*, 2020). As known signals become better characterized, filters can be used for the detection of known objects within noisy backgrounds.

The characterization problem will be aimed at recovering the acoustic source properties of the aerial object itself, that is, identifying the intrinsic and extrinsic mechanisms responsible for the production of sound at or near the object. Analytical solutions will need to use the time and frequency domain characteristics of a signal received by a sensor on the ground and deconstruct the effects of the transmission path. This requires a good understanding of the distance traveled and the atmospheric characteristics along the sound path. We will explore solutions by using characteristics of signals from known aircraft and measuring how the transmission path modifies them. ADS-B data identifies individual aircraft, which enables us to record the same aircraft multiple times under varying conditions. In addition, we can use data from sites where the same aircraft routinely fly overhead to examine the variations of sound with distance, bearing, and altitude. In this way we can begin to understand the transformation of sound from source to receiver in the local environment of each field site.

## 6.2. *Audible band analyses*

As discussed in Sec. 6.1, general signal processing methods for the three frequency bands captured by AMOS will incorporate similar elements, however, specific analytical techniques will likely differ between bands. Here, we focus on the audible band recorded with AudS. In Phase 1, we will collect a large set of data and use it to explore the capabilities of the recording systems; correlate signals with aircraft ADS-B flight data for the labeled database; understand characteristic structures in the data; and learn about the effects of ambient noise, acoustic environment, and transmission pathways on the source signal. In this section, we review a set of preliminary raw data collected by the audible microphone system, which collects sound from 10 Hz to 20 kHz and is sampled at 44.1 kHz after passing through a 22 kHz low-pass filter.

### 6.2.1. *Example aircraft spectrogram*

Preliminary recordings of aircraft were collected at the Center for Astrophysics (CfA) by mounting the audible microphone on a roof railing. Data sampled at 44.1 kHz on May 31, 2022 recorded sound from an Airbus 319 passenger jet that passed overhead at a height of 1.2 km as determined by the ADS-B transponder record. A windowed Fast Fourier Transform (FFT) spectrogram (8192 points, 50% overlap) of the nearly one-minute time series reveals the rich and complex information contained in audible data (Fig. 8). Lighter color represents stronger signals in this plot of signal power as a function of time and frequency. Three prominent features of the aircraft sound can be seen: Doppler shift, destructive interference, and acoustic shadow zones.

The 320 class of Airbus, of which the 319 is a member, is known to produce a whistle due to resonance in their fuel overpressure protector cavities (CDA, 2022). This whistle is typically around 500–600 Hz and shows up in this spectrogram of captured sound as tonal components (Berckmans *et al.*, 2006). This signal, along with several harmonics, exhibit classic Doppler shifted S-curves. The frequency at time zero is up-shifted as the aircraft approaches from afar, falls to near its true value at an inflection point as the aircraft passes overhead at time = 20 s, and then continues to fall as the plane recedes (as noted on Fig. 8). This change in frequency can be used to estimate a minimum bound on aircraft velocity (Tong *et al.*, 2013).

From 10 s to 35 s, a stack of "U" shaped curves is apparent, with higher frequencies visible as the aircraft closes in on the receiver from 15 s to 26 s. These bands of dark and light are due to a reflected signal from the ground creating destructive interference with the direct signal coming in from above. The "U" shape is due to the frequency dependence of destructive interference on the changing angle of incidence (Tong *et al.*, 2013).

As one moves across this spectrogram in time, there are vertical bands where the power level of the signal waxes and wanes through the entire bandwidth of frequencies. For example, from 26 s to 29 s, an acoustic shadow zone appears. This feature may be due to inhomogeneities in the density of air, or in variations in wind speed and direction, along the acoustic transmission path. These disturbances can result from processes such as thermal layering and turbulent eddies.

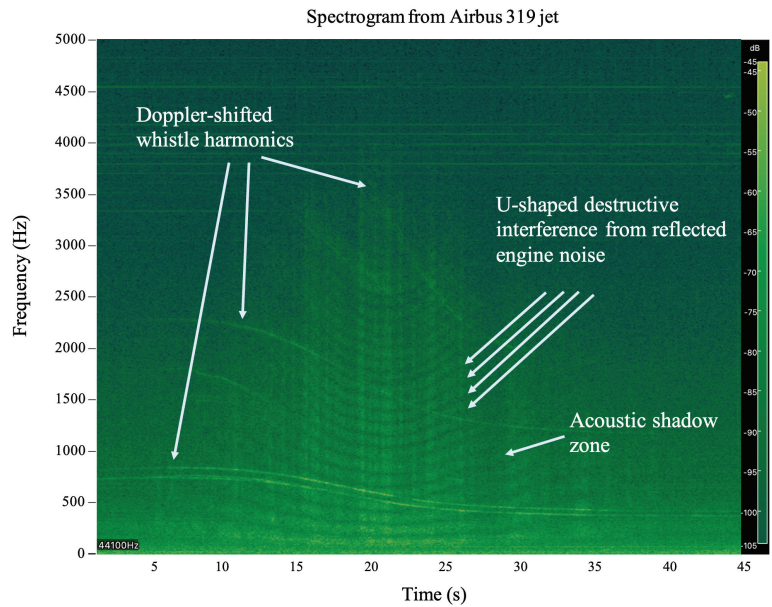

Fig. 8.   (Color online) Time-frequency spectrogram of audible sound from an Airbus 319 passenger jet passing at a height of 1.2 km over the AMOS system on the CfA roof at Harvard University. Relative power in dB scale is on the left; lighter shades represent higher relative dB and thus higher SPL. Doppler shifted frequencies, destructive interference, and acoustic shadow zones are visible.

### 6.2.2. *Database generation*

As a first step in our investigation of analysis methods, we created a labeled database of audible sound recordings of both the ambient noise environment and of periods when aerial vehicles could be detected by ear. The data had been previously recorded by the AudS while mounted on the roof of the CfA continuously for five days. The data were reviewed and segmented by hand into 60-second wav file clips where an airplane could be heard. These sound recordings were then time-correlated with ADS-B flight recorder data in order to provide information on the location, speed, altitude and ID of any airplanes or helicopters in the area during the sample window. Samples of ambient background noise were reviewed and randomly segmented into 60-second wav file clips from a larger ambient sample where no airplanes or helicopters were heard. A simple two-dimensional Pandas DataFrame, or table, was then created, containing timestamps, vehicle signal class (e.g. Airbus 319-132), filename, and the file locations containing the represented data captures. The two most common vehicle signal classes in our test sample of 50 aircraft were the Cessna-402C (11) and Airbus A321-231 (4).

### 6.2.3. *Data processing*

Initial data processing methodology has three main components: Selection of the raw audio files;

application of a moving Short-Time Fourier Transform (STFT) to the times series to generate a spectrogram showing power as a function of frequency and time; and generation of the time-averaged spectrum for the entire clip, showing characteristic power versus frequency. We employ the pydub (jiaaro, 2022) (for .wav files) and librosa (McFee *et al.*, 2022) (for audio analysis) python packages, which allow us to transform the time series data into the frequency domain. Exploring acoustic time series in the frequency domain can reveal power at characteristic frequencies such as those generated by specific jet engines, propeller designs, fuselages, or aircraft types.

### 6.2.4. *Feature extraction*

Focusing on the audible data, there are several analyses that could be developed to extract meaningful features from signals in the time-frequency domain. The following methods will be explored in future work:

- Background subtraction and key signature detection: there are a range of techniques to perform background subtraction (Xie *et al.*, 2016; Bouwmans *et al.*, 2019), which would allow us to identify key frequencies in the spectra using traditional methods and could be developed with smaller data samples.

- Extracting cepstral coefficients: these represent the detection of periodicity in the frequency spectrum (e.g. harmonics of 100 Hz would generate a single cepstral coefficient) and are often used in the analysis of human speech (Beigi, 2011). This would allow the reduction of the data into a smaller parameter space for rapid machine learning techniques.
- Generating time series-based neural network models: these are commonly used in learning order dependencies in sequence prediction problems, but require large data sets to generate a predictive model (Lezhenin *et al.*, 2019).
- Exploring combined and alternative analyses: additional methods may be required to sufficiently characterize unique ambient environments and source characteristics in the audible frequency band (Sharma *et al.*, 2020).

## 7. Conclusions

The AMOS acoustic sensor and data capture suite, with its single microphones in the infrasonic, audible, and ultrasonic bands, can operate continuously in the field, capturing at ground level the acoustic signals from a wide range of aerial objects. The distance to detectable objects from the suite's location will depend on the frequencies of interest, with lower frequencies having longer detection distance than higher frequencies, and sound source power levels. This detection distance will also depend on the ambient noise environment in each frequency band of interest.

The information collected from acoustic signals through the use of feature extraction techniques in the time and frequency domains, machine learning, and labeled databases of known phenomena will assist the Galileo Project in identifying aerial phenomena.

With proper instrument calibration and ambient background measurements, coupled with information about the propagation path of the sound, time and frequency domain analysis of the data may be used to characterize intrinsic and extrinsic sound generation mechanisms of aerial phenomena.

Calibration of the system, including ongoing correlations of signals with known aircraft flight data, monitoring the ambient noise environment, and understanding the sound transmission paths, will be critical for determining the decibel buffer level (Sec. 5.2) for triggering a detection. The choice of buffer level will, in turn, inform the level of confidence with which aerial objects are detected, identified, and classified by the acoustic system.

Future work will include developing sensor arrays in infrasonic and audible bandwidths. This will give us the ability to localize and track moving aerial objects. It will provide an opportunity to use signal processing techniques for noise reduction, beamforming, and distinguishing sounds simultaneously emitted from sources at different locations. The use of multi-sensor arrays will increase our detection distance and allow better characterization of the received acoustic signals and the corresponding sound generating mechanisms of aerial objects.